\documentstyle{aipproc}

\begin{document}
\title{Solving the QCD Hamiltonian for bound states}

\author{Elena Gubankova, Chueng-Ryong Ji, and Stephen R. Cotanch}
\address{Department of Physics, North Carolina State University, 
Raleigh, NC 27696-8202}

\maketitle

\begin{abstract} The systematic approach to study bound states in
quantum chromodynamics is presented. 
The method utilizes nonperturbative flow equations  
in the confining background, that makes possible 
to perform perturbative renormalization and 
to bring the QCD Hamiltonian to a block-diagonal form with 
the number of quasiparticles conserving in each block. 
The effective block-diagonal Hamiltonian provides 
constituent description for hadron observables.
The renormalized to the second order effective Hamiltonian 
of gluodynamics in the Coulomb gauge is obtained at low energies. 
The masses for scalar and pseudoscalar glueballs are predicted. 
\end{abstract}

\section{Introduction}

One of the most difficult and less understood problems
in quantum chromodynamics is the treatment of the bound state systems.
There are different sources of difficulties.
For example, it is quite a common observation known in the spectroscopy,
that the splitting between the vector mesons does not depend
on flavor, say
\begin{equation}
m(\rho^{\prime})-m(\rho) \sim m(\psi^{\prime})-m(J/\psi)
\,\label{}\end{equation}
which is true experimentally. This fact can not be explained
in terms of the canonical QCD interaction, which is given
essentially by the strong coupling constant. 
Indeed, if expressed in terms of an invariant mass $s$,
Eq.(1) implies that the $J/\psi$ is dual to a much larger interval of $s$
than the $\rho$ because the $c$-quark is heavy. However, 
the coupling constant runs as a function of an invariant mass, $\alpha(s)$, 
and is flavor blind, thus the canonical QCD interaction
should be much weaker for the $J/\psi$ than for the $\rho$.
This suggests that something is missing when described only
in terms of the perturbation theory. The strong coupling constant alone 
does not provide for strong interactions being {\it strong}.

Consider the scaling of Quantum Chromodynamics
from high to low energies. In the ultraviolet region  
(at the bare cutoff scale $\Lambda\rightarrow\infty$)
the strong interactions are given by canonical QCD, 
which is conformally invariant, in particular 
this means scaling invariance (there is no scale in the theory) 
in the chiral limit. Moreover a perturbative treatment is possible
due to asymptotic freedom. In asymptotic free theories (QCD) 
the coupling constant grows at low-energies and gets strong, 
that stops the asymptotic freedom 
at some moderate scale $\Lambda_0\sim\Lambda_{QCD}$.
This scale appears in the theory when the perturbative renormalization
of the coupling constant is performed, and the $\Lambda_0$ 
is the Landau pole in the effective running coupling constant
provided the renormalization group invariance.
This is called dimensional transmutation,
when scaling invariance breaks through the renormalization.
Then experiment tells us that this is not the only scale in the theory.
There are at least two more characteristic scales in the hadron physics.
The mass gap of the hadron bound state, say, given by
the square root of the string tension, where the nonperturbative 
phenomenon of confinement takes place and the bound states 
of quarks and gluons form. The scalar $\pi$-meson
sets up a scale of chiral symmetry breaking -
another phenomenon of nonperturbative physics.
The scales are displayed as
$m(\pi)\ll\sqrt{\sigma}\ll\Lambda_{QCD}$.
 
{\it To summarize:} 
Knowing perturbation theory 
alone is not enough to describe bound states in hadronic physics.\\  

Other difficulties in the QCD bound state problem 
are of a more general kind -
relativistic nature of quantum field theory.
First, the number of particles in any state is not fixed
because of particle creation and annihilation in vacuum.
Second, there are states with (infinitely) large energies.
An attempt to treat both problems was done in \cite{GuWe}
by using the method of flow equations. The idea is to find
a unitary transformation that transforms the Hamiltonian operator
to a block-diagonal form, where each block conserves 
the number of particles. The Hamiltonian matrix can be represented
in the particle number space as
\begin{eqnarray}
   H = \left(
   \begin{array}{cc}
      PHP & PHQ \\
      QHP & QHQ
   \end{array}\right)
\end{eqnarray}  
where $P$ and $Q$ are projection operators on the subspaces
with different particle number content.
The flow equations \cite{Wegner} bring the Hamiltonian matrix Eq.(2)
to the form
\begin{eqnarray}
   H_{\rm eff} =\left( 
   \begin{array}{cc}
     PH_{\rm eff}P & 0            \\
     0             & QH_{\rm eff}Q
   \end{array}\right)
\end{eqnarray} 
where the two blocks of the effective Hamiltonian 
decouple from each other. It may be simpler then to solve 
for bound states within one block, say $PH_{eff}P$,
than to diagonalize the complete Hamiltonian, Eq.(2),
of the original problem. Since, generally, the number of particles
in $P$ and $Q$ spaces is arbitrary, one can reduce in this way
the bound state problem with many particles
to a few body problem.

It turns out that the second question on possible presence of 
the ultraviolet divergences is solved also 
by the method of flow equations. 
Flow equations perform a set of (infinitesimal small) 
unitary transformations, where the flow parameter,
which controls the transformation, has the dimension
of the inverse of the energy square, $l\sim 1/E^2$.
Therefore by using flow equations to block-diagonalize   
the Hamiltonian one eliminates the particle number changing contributions  
not in one step but rather continuous for the different energy differences 
in sequence. This procedure enables one to separate 
the ultraviolet divergent contributions and to find
the counterterms associated with these divergences. 
This covers the UV-renormalization for Hamiltonians \cite{GlWi}.

{\it To summarize:}
More generally, flow equations perform Hamiltonian renormalization
in the ``particle number'' and in the ``energy'' spaces
in the sense that the effects of the high Fock states and
the effects of the large energies, respectively, are encoded
in the effective low-energy Hamiltonian, 
which operates in the space of the few low Fock components.\\   

This program was applied quite successfully to QED 
to calculate the positronium spectrum \cite{GuWe}. 
The key is the validity of the perturbation theory    
in the bare coupling constant for the characteristic energy scale 
of positronium bound state.
Obviously, it is a bad idea to apply naively the same scheme for QCD.
It is not possible to find the fixed number representation 
for the Hamiltonian in the case of {\it strong} interactions,
where one does not have any control over the process of 
the creation and the annihilation in vacuum
of bare quarks and gluons with {\it small} current masses. 
In the language of flow equations convergence
can not be achieved when calculated in terms of bare parameters.

The way out is suggested by nature itself.
One should consider ``{\it confined} QCD''.    
By using flow equations one constructs then 
the effective QCD Hamiltonian $H_{eff}(q,g)$,
where current quarks and gluons 
acquire masses of the order of $m_{constituent}\sim 1 GeV$
and become constituent degrees of freedom (see below).               
The value of the constituent mass plays the role of the energy gap
between the sectors in $H_{eff}$.
Schematically, the block-diagonal effective Hamiltonian
has the form
\begin{eqnarray}
   H_{\rm eff} =\left( 
   \begin{array}{c|c|c}
     q\bar{q} &  &        \\  \hline   
         & q\bar{q}g &      \\ \hline 
      &  &  q\bar{q}q\bar{q}     \\
       &  &  gg 
   \end{array}\right)
\end{eqnarray} 
where $q$ and $g$ are constituent quarks and gluons, respectively;
empty cells denote zero, 
to the order calculations are done, matrix elements. To this order
the different sectors of the effective Hamiltonian, Eq.(4), 
describe approximately, when going down, 
the bound states of mesons, hybrids, glueballs.
Actually, such a description with the fixed number of constituents 
is quasiclassical and nonrelativistic, and is known
from the constituent quark model. 
Physically, the picture is the following.
The strong confining interaction, acting inside  
each ``diagonal'' (particle number conserving) sector, 
produces heavy gluon (quark) 
only in the small volume -- in the ``bag''. 
No free propagating heavy gluons (quarks) are produced.
Therefore the ``bags'' do not interact with each other, 
decoupling in $H_{eff}$ and approximating the hadron bound states. 

The matrix elements of the ``off-diagonal'' (particle number changing) 
sectors are governed by the canonical interaction -- typically
by the Coulomb term of the strength equal to the inverse 
of Bohr radius or the current quark mass, $m_{current}$
is of the order of several $MeV$. 
In the presence of the strong confining interaction
in the ``diagonal'' sectors the mixing between the sectors
is strongly suppressed. One can introduce a small parameter, say, 
\begin{equation}
\alpha_s\frac{V_{12}}{E_1-E_2}\sim 
1\cdot\frac{m_{current}}{m_{constituent}}
\sim 0.1-0.01
\\,\end{equation}    
where $V_{12}$ is the Coulomb interaction, and $E_1-E_2$ is 
the energy difference between the first and the second ``diagonal'' sectors.
Perturbation theory with respect to the small parameter, 
Eq.(5), holds between the sectors (but not inside the sector
where the confining interaction is strong). By applying flow equations 
to block-diagonalize the Hamiltonian one gets to leading order 
a closed chain of decoupled equations, which can be solved analytically.  
The whole is true provided there is a strong confining interaction
in the ``diagonal'' sectors.

{\it To summarize:} In the theory of strong interactions   
confinement is important
to provide the bound states. In the present approach 
confinement makes it possible to bring, by flow equations,
the QCD Hamiltonian to a block-diagonal form 
with a fixed number of quasiparticles in each sector. 
The elementary degrees of freedom (quasiparticles) 
become constituent quarks and gluons, which acquire masses of order $1 GeV$.
The block-diagonal effective Hamiltonian approximately describes then 
the different hadronic bound states.\\

The main idea of the approach is to find the representation
for QCD Hamiltonian with the fixed number of quasiparticles,
where the sectors with different particle number content
decouple from each other. There can be some special cases
when one should take into account the mixing between the sectors. 
In other words the physical state is not given by the pure component
of the composite system. 
The mixing between the high excited state from
the previous sector and the ground state from the next sector
of the effective Hamiltonian may be possible
(for example, the mixing between some excited meson 
and the low lying hybrid state).
In systems with light quarks the influence of coupled channels 
can be essential. In the strongly coupled effective meson models 
one includes the effects from the coupled channels directly
by mixing the scalar and pseudoscalar channels
($q\bar{q}$ and $q\bar{q}q\bar{q}$).
The effect is about $50$ percent.

Special consideration is required in the case of the light quarks,
where chiral symmetry breaking (CSB) is important. 
The present approach includes confinement 
and is like the ``bag model'' or the ``constituent quark model'',
but it does not include CSB. By implementing CSB in this picture, 
the scalar $\pi$-meson can be viewed as a bound state of the two 
constituent quarks and simultaneously manifests the Goldstone nature.

{\it Motivation:}
In order to disentangle the both problems of confinement
and CSB we consider the pure gluodynamics 
(see the next section) \cite{GuSwJiCo}. 
The motivation for this study is to set up a kind of 
a constituent gluon model, with the confining interaction imposed, 
to describe glueball bound states.\\

The specific difficulty of QCD is that the canonical QCD Lagrangian 
does not manifest explicitly confinement. 
As far as the mechanism of confinement is concerned
one can proceed along several ways.
To reveal confinement one uses the suitable formulation of QCD:
lattice form, or the special choice of the gauge fixing 
(for example, maximal Abelian projection). Another option is to study 
other than QCD theories, but that have the same infrared behavior 
as QCD and are confined: Super Yang Mills theory, some toy gauge models 
(for example, Abelian Higgs Model). There may appear some unphysical
degrees of freedom in these theories.
If one is not interested in the mechanism of confinement,
one includes the latter explicitly into QCD. The simplest way is to use 
the potential model, successfully tested in phenomenology, 
where the potential between the color charges 
is given by a sum of Coulomb and confining interactions.
This suggest the definite choice of the gauge for the Hamiltonian.
We work in the Coulomb gauge, where the Coulomb interaction arises
from the gauge fixing procedure. We add then confinement
to be able to block-diagonalize to the effective Hamiltonian, which 
describes the bound states.

The Coulomb gauge is the natural gauge to get the constituent 
hadron picture. The Coulomb interaction appears there not as a perturbative 
(propagating in time) one gluon exchange, but rather as a solution 
of the gauge fixing constrains. Therefore the Coulomb term
describes an instantaneous interaction, which is consistent
with the nonpropagating massive gluon arising in our approach.
Note, that the massive gluon mode arises only in the presence 
of confinement and confinement sets the scale for the gluon mass.
The obvious drawback of the approach is the violation of 
the gauge invariance by the massive gauge fields.

{\it To summarize:}
The Coulomb gauge has an appealing property of
the simple extension of the model to the confining case
and is consistent with the constituent picture for hadrons.

\section{Low-energy gluodynamics in the Coulomb gauge}

As noted above we applied the method to pure gluodynamics 
\cite{GuSwJiCo}.
In this section we outline the strategy of this study.

1. {\it QCD $H_{can}$.}\\ 
The starting point is the canonical QCD Hamiltonian 
(pure gluodynamics) in the Coulomb gauge
$\nabla\cdot{\bf A}=0$: $H_{can}({\bf A},{\bf \Pi})$,
where physical degrees of freedom are the transverse gauge 
fields ${\bf A}$ and their conjugate transverse momenta 
${\bf \Pi}$.

2. {\it $H_{can}=H(g=0)+O(g)+O(g^2)$.}\\
We expand the canonical QCD Hamiltonian in the Coulomb gauge 
perturbatively to the second order in the bare coupling constant.
Then to the leading order the Faddeev-Popov determinant can be 
approximated by unity, that reduces
the instantaneous term, arising from the gauge fixing, 
to the pure Coulomb interaction.

3. {\it Current (perturbative) basis and 
the trivial vacuum $|0\rangle$.}\\
We choose the trivial (perturbative) vacuum $|0\rangle$  
and construct the perturbative basis of free ({\it current}) particles:
$a^{\dagger}({\bf k})|0\rangle$ creates one (perturbative) gluon 
with zero mass, i.e. the gluon energy $\omega_{\bf k}=|{\bf k}|$, 
etc., and the vacuum is defined as $a|0\rangle=0$.

We express the canonical QCD Hamiltonian (section 2) 
in this perturbative Fock space,
and normal order the result with respect 
to the trivial vacuum state $|0\rangle$.
Denote the normal ordered canonical QCD Hamiltonian as 
${\bf :}H_{can}{\bf :}$.

4. {\it Regularization and perturbative renormalization 
(scheme).}\\ 
The normal ordered Hamiltonian 
${\bf :}H_{can}{\bf :}$ contains ultraviolet (UV) divergent terms
(UV-divergent loop integrals).
We regulate UV-divergences by the cutoff function $f(q,\Lambda)$ 
(the explicit form of the cutoff function is specified further).
This is the first time when we have introduced an energy scale 
in the theory -- the bare cutoff $\Lambda\rightarrow\infty$.
To remove the cutoff sensitivity we renormalize the Hamiltonian 
by adding the counterterms associated with these divergences.
Schematically, the renormalized Hamiltonian is written as 
$H_{ren}(\Lambda)= H_{can} + \delta X_{CT}(\Lambda)$,
where $\delta X_{CT}(\Lambda)$ is a set 
of (unknown) counterterm operators, which we define further
\footnote{In the given (perturbative) basis this equation 
reads
${\bf :}H_{ren}(\Lambda){\bf :}=
{\bf :}H_{can}{\bf :}+{\bf :}\delta X_{CT}(\Lambda){\bf :}$,
where $"{\bf :}"$ stands for normal ordering in the (perturbative) vacuum
}. 

5. {\it Flow equations perturbatively.}\\ 
To find the explicit form of the counterterms    
and to scale down the Hamiltonian we run flow equations perturbatively.
Also the form of the cutoff (regulating) function is specified 
by flow equations. Generally, flow equations define
the prescription of regularization and make possible 
to perform the perturbative renormalization
\footnote{
Flow equations perform a set of unitary transformations
to block-diagonalize the Hamiltonian
$H(l,l_0)=U^{-1}(l,l_0)H(l_0)U(l,l_0)$, where $l$ is the flow parameter
with the connection to the energy scale $l=1/\lambda^2$,
$l_0$ is the initial value corresponding to the bare cutoff $\Lambda$
introduced by the regularization before (section 4).}.
Technically, since the Hamiltonian depends on the cutoff scale 
through the flow parameter,
one finds in the given order of perturbation theory (PT) 
the divergent part of the difference
between the Hamiltonian operators given at the two scales, say
$(H(\Lambda_2)-H(\Lambda_1))$ with 
$\Lambda_{QCD}\ll\Lambda_2\le\Lambda_1\le\Lambda$. 
One absorbs then these divergences in the counterterms -- 
local operators with the symmetries of the canonical Hamiltonian, 
to provide the renormalization group invariance
(called in the context of Hamiltonian renormalization
"coupling coherence" \cite{GlWi}). 
This completes the procedure of renormalization, performed 
by flow equations, to this order. 
One can proceed in this way order by order in PT 
to find (all) the counterterms systematically.  
Note, that it is enough to find the gradient of the Hamiltonian
in the energy space to define the counterterms.
Renormalization group invariance (RGI) insures, that
the renormalized Hamiltonian preserves the form
of the (original) canonical Hamiltonian, but only the coupling constants 
and the mass operators (that are usually classified as relevant and 
marginal operators in renormalization group sense)
start to run with the cutoff scale.
(We do not consider here, at least in the few lowest orders of PT, 
possible irrelevant operators, that may cause new type of divergences 
than are carried by coupling constants and masses).

Using flow equations we run the effective Hamiltonian 
downwards from the bare cutoff $\Lambda$ to some intermediate 
scale $\Lambda_0\sim\Lambda_{QCD}$, where perturbation theory 
breaks down. Due to the RGI the ``physical gluon'' stays massless 
through this perturbative scaling. We can not proceed 
with flow equations perturbatively further.
The result of this stage is the renormalized 
(to the second order of PT \cite{GuSwJiCo}), 
effective Hamiltonian, defined at some compositness scale $\Lambda_0$: 
${\bf :}H_{ren}(\Lambda,\Lambda_0){\bf :}$ with $\Lambda_0\sim\Lambda_{QCD}$
and bare cutoff $\Lambda\rightarrow\infty$,  
and semicolon means normal-ordering 
in the trivial vacuum $|0\rangle$.
Though the renormalized Hamiltonian is obtained in the perturbative
frame, it can be represented (regardless of the Fock basis) 
in terms of the fields ${\bf A}$ and ${\bf \Pi}$ \cite{GuSwJiCo}.
(It is a consequence of the RGI).
We denote the resulting renormalized Hamiltonian at the scale $\Lambda_0$
as $H_{ren}(\Lambda,\Lambda_0)$.

6. {\it Confinement.}\\ 
We introduce confinement as a linear rising potential,
that enables to run flow equations ``nonperturbatively''
(see introduction) until complete diagonalization of the Hamiltonian.
In the renormalization group sense this ``spoils'' the theory: 
there arises the massive gluon mode. 
But the presence of confinement
is necessary to find the representation with a fixed number of quasiparticles
(constituent massive gluons) for the effective Hamiltonian,
which provides the constituent picture (see introduction).
Confinement (string tension) sets the (hadron) scale for the gluon mass.

The instantaneous interaction contains two pieces, 
the sum of the Coulomb and confining potentials.
Denote the renormalized effective Hamiltonian with 
confinement embedded as $H_{eff}(\Lambda,\Lambda_0)$.

7. {\it Constituent (nonperturbative) basis and 
the QCD vacuum $|\Omega\rangle$.}\\ 
As far as confinement is introduced the trivial vacuum $|0\rangle$  
and the perturbative basis of free (current) particles,
$\omega_{\bf k}=|{\bf k}|$, define no longer the minimum ground state.
Therefore, we introduce the (arbitrary) basis,    
where the gluon energy $\omega_{\bf k}$ is kept unknown,
and is defined further variationally.
Correspondingly, the (nontrivial) QCD vacuum $|\Omega\rangle$ 
is defined as $\alpha|\Omega\rangle =0$, and the Fock space 
of {\it constituent} particles is given:
$\alpha^{\dagger}|\Omega\rangle$
creates the quasiparticle with the energy $\omega_{\bf k}$, etc. 
\footnote{
The change of basis  
from the (perturbative) current, $\omega_{\bf k}=|{\bf k}|$, 
to the (nonperturbative) constituent, with some $\omega_{\bf k}$, 
can be written as Bogoluibov-Valatin (BV) transformation  
from the ``old'', $a,a^{\dagger}$, to the ``new'', $\alpha,\alpha^{\dagger}$, 
operators: $a_{\bf k}={\rm ch}\phi_{\bf k}\alpha_{\bf k}+
{\rm sh}\phi_{\bf k}\alpha_{-{\bf k}}^{\dagger}$ 
with BV angle $\phi_{\bf k}$ given by
${\rm ch}\phi_{\bf k}=1/2(\sqrt{k/\omega_{\bf k}}+
\sqrt{\omega_{\bf k}/k})$. The connection between 
the ``old'', $|0\rangle$, and the ``new'', $|\Omega\rangle$, vacuum states 
is given  $|\Omega\rangle={\rm exp}\left( 
\frac{1}{2}\sum_k{\rm th}\phi_{\bf k} 
a^{\dagger}_{\bf k}a^{\dagger}_{-{\bf k}}\right) |0\rangle$.
It was used in the work \cite{SzSwJiCo} to transform the QCD Hamiltonian
into the constituent basis.
}. 
The renormalized effective Hamiltonian $H_{eff}(\Lambda,\Lambda_0)$
at the scale $\Lambda_0$ (section 6), written through the physical fields
${\bf A}$ and ${\bf \Pi}$ and having confinement, 
is decomposed in the trial (constituent) basis and normal-ordered
with respect to QCD vacuum $|\Omega\rangle$. The unknown gluon energy
is variational parameter in the calculations.
We combine the terms in the effective Hamiltonian 
in each particle number sector according
to the power of coupling constant $O(g^n)$ ($n=0,1,2$)
\footnote{The higher order terms in the effective Hamiltonian
are suppressed by the inverse powers of (heavy) gluon mass,
which is of order of hadron scale (see introduction).}.
In the absence of confinement the effective Hamiltonian 
preserves the form of canonical Hamiltonian due to RGI,
with the proper change $|{\bf k}|\rightarrow\omega_{\bf k}$.
In the presence of confinement the canonical form is violated by 
the second order terms in the effective Hamiltonian, 
which contribute higher orders $O(g^3), etc.$ in flow equations.

We aim to find the effective Hamiltonian
after the scaling downwards from $\Lambda_0$
to a hadron scale, say $\sqrt{\sigma}$.
Since the effective Hamiltonian preserves
the canonical form at least to the second order,
the ``perturbative'' terms obtained by flow equations in 
section 5 match the ``nonperturbative'' terms arising 
when applied flow equations to $H_{eff}(\Lambda,\Lambda_0)$.

We denote the effective Hamiltonian 
in constituent basis as 
${\bf ::}H_{eff}(\Lambda,\Lambda_0){\bf ::}$,
where $''{\bf ::}''$ stands for normal-ordering in the QCD vacuum. 

8. {\it Flow equations in the confining background.}\\ 
We run flow equations in the confining background 
to block-diagonalize 
the effective Hamiltonian ${\bf ::}H_{eff}(\Lambda,\Lambda_0){\bf ::}$ in 
the nonperturbative basis and to find consistently 
all the terms to the second order.  
Free Hamiltonian and confining interaction are included 
in ``diagonal'' sector,
the triple-gluon vertex forms ``nondiagonal'' sectors,
that should be eliminated.
We bring the Hamiltonian ${\bf ::}H_{eff}(\Lambda,\Lambda_0){\bf ::}$ 
to a block-diagonal form, 
where diagonal blocks decouple from each other including the second order.
The leading UV-behavior of the arising to the second order terms
is cancelled by the mass counterterm. Generally,
this approach allows to include perturbative QCD corrections
into nonperturbative calculations of many-body techniques.
The resulting block-diagonal effective, renormalized Hamiltonian
is given at hadronic scale, ${\bf ::}H_{eff}(\Lambda,\sqrt{\sigma}){\bf ::}$. 
For simplicity we denote it as $H_{eff}$.

9. {\it Gap equation (variational calculations).}\\
The requirement of block-diagonal form does not fix 
the effective Hamiltonian completely. The remaining freedom
to unitary transform inside of each block is fixed 
by minimizing the ground state (the vacuum expectation value of
the effective Hamiltonian with respect to QCD vacuum)
\begin{eqnarray}
d \langle\Omega|H_{eff}|\Omega\rangle/d\omega =0
\end{eqnarray}
to find the trial gluon energy $\omega({\bf k})$.
The variational function $\omega({\bf k})$ is defined.
As a result the gluon acquires a nonzero mass, $m\sim 0.5 GeV$,
and an iterative procedure of flow equations is performed
with respect to the small parameter $1/m$.
The next following sector in the effective Hamiltonian
is suppressed by this factor, that provides
the convergence for the flow equations (see introduction).

10.{\it Solving for $H_{eff}$.}\\
There are two parameters in the method, the two scales:
$\Lambda_0$, which defines the counterterms
and regulates the perturbative radiative corrections
to the effective Hamiltonian $H_{eff}$,
and $\sqrt{\sigma}$, where $\sigma$ is the string tension 
defining the nonperturbative confining potential.

Since the effective Hamiltonian is block-diagonal,   
one can solve for the bound states in any interesting sector 
(actually in the few lowest sectors).
We solve for the glueball bound state in the two-body sector.
The result is the glueball spectrum.

\section{Summary}
1. {\it Renormalization}\\
Renormalization was performed to the second order.
We combined the individual counterterms in one- and zero-body sectors
to the resulting mass counterterm, written
in the field representation (independent of the basis)
$\delta X_{CT}(\Lambda) = m^2{\rm Tr}\int d{\bf x}{\bf A}^2({\bf x})$
with 
$m^2 = -\frac{\alpha_s}{\pi}N_c\frac{11}{12}\Lambda^2$
Remarkably, when the quark sector is added in the same fashion, 
the algebraic coefficient in the propagator correction reproduces
the QCD $\beta$-function. 
This particular feature of the Coulomb gauge 
supports our regularization prescription, which 
follows from flow equations.

2. {\it Glueball}\\
We specify the two parameters: 
the string tension is defined by the lattice calculations 
$\sigma=0.2GeV^2$;
the cutoff $\Lambda_0$ is found from the gluon condensate
to agree with the result of the sum rules,
the condensate term is obtained 
$\langle G^2\rangle\sim 1.3\cdot 10^{-2}GeV^4$
with $\Lambda_0=4GeV$.

The solution of the gap equation can be parameterized as
$\omega({\bf k})=k+m(0){\rm exp}(-k/\kappa)$,
where the effective gluon mass is obtained 
$m(0)=0.90GeV$ and $\kappa=0.95GeV$.

The glueball mass spectrum in scalar and pseudoscalar channels
is given in Tamm-Dancoff approximation in \cite{GuSwJiCo}. Roughly,
the mass of the lowest scalar glueball $0^{++}$, $1760 MeV$,
is twice of the effective gluon mass $m(0)$.

\vspace{1cm}
{\bf Acknowledgments.} One of the authors (E.G.) is thankful
to Karen Avetovich Ter-Martirosyan for useful discussions
and introducing the idea.


\begin{references}
\bibitem{Wegner} F. Wegner, Ann. Phys. {\bf 3}, 77 (1994).
\bibitem{GlWi} St. D. Glazek and K. G. Wilson, Phys. Rev. {\bf D 48},
5863 (1993); {\it ibid.} {\bf 49}, 4214 (1994).
\bibitem{GuWe} E. Gubankova, F. Wegner, Phys. Rev. {\bf D 58},
025012 (1998), hep-th/9710233.
\bibitem{SzSwJiCo} A. Szczepaniak, E. S. Swanson, C.-R.Ji, and
S. R. Cotanch, Phys. Rev. Lett. {\bf 76}, 2011 (1996).
\bibitem{GuSwJiCo} E. Gubankova, E. Swanson, Ch.-R. Ji, S. Cotanch,
hep-ph/9905527. 
\end{references}
\end{document}